\documentclass[twocolumn,showpacs,preprintnumbers,amsmath,amssymb,prl]{revtex4-1}
\usepackage{graphicx}
\usepackage{dcolumn}
\usepackage{bm}

\begin{document}

\title{Intrinsic giant Stark effect of boron-carbon-nitride nanoribbons
with zigzag edges}

\author{T. Kaneko}
 \affiliation{Nanoscale Theory Group, NRI, AIST,
1-1-1 Umezono, Tsukuba, Ibaraki 305-8568, Japan}

\author{K. Harigaya}
 \affiliation{Nanoscale Theory Group, NRI, AIST,
1-1-1 Umezono, Tsukuba, Ibaraki 305-8568, Japan}

\author{H. Imamura}
 \email{h-imamura@aist.go.jp}
 \affiliation{Nanoscale Theory Group, NRI, AIST,
1-1-1 Umezono, Tsukuba, Ibaraki 305-8568, Japan}

\date{\today}

\begin{abstract}
 Electronic properties of zigzag boron-carbon-nitride (BCN) nanoribbons, where the outermost C
 atoms on the edges of graphene nanoribbons are replaced by B or N atoms,  
 are theoretically studied using the first-principles calculations.
 We show that BCN nanoribbons are metallic,
 since several bands cross the Fermi level.
 For BCN nanoribbons in a rich H$_2$ environment,
 the so-called nearly free electron state appears just above the Fermi level because of the intrinsic giant Stark effect 
 due to the internal electric field of a transverse dipole moment.
 The position of the nearly free electron state can be controlled by 
 applying an electric field parallel to the dipole moment.
 The hydrogenation of the nitrogen atom is necessary
 for the appearance of the giant Stark effect in BCN nanoribbons.
 We also discuss the effect of stacking order on the intrinsic giant
 Stark effect in bilayer BCN nanoribbons.
\end{abstract}


\maketitle

Graphene nanoribbon (GNR) is a nanometer-wide strip cut from a 
graphene (Gr) sheet on which C atoms are arranged in a honeycomb lattice.
The electronic and magnetic properties of GNR have been studied intensively
 by several groups
\cite{Fujita1996jpsj,Nakada1996prb,Wakabayashi1999prb,Miyamoto1999prb}, which predicted the presence of
a peculiar electronic state at zigzag edges, which is called the edge state.
GNR can be fabricated by the use of electron-beam lithography
\cite{Han2007prl}, the unzipping of carbon nanotubes (CNTs)
\cite{Jiao2009nature,Kosynkin2009nature}, and the use of bottom-up processes
\cite{Cai2010nature}. The presence of edge states of GNR were
confirmed experimentally by Tao {\it et al.} \cite{Tao2011nphys}.

One of the most interesting properties of Gr-based nanomaterials, such as
CNT and GNR, is the controllability of energy band structures using an
applied electric field, i.e., the giant Stark effect \cite{O'Keefe2002apl,Son2005prl,Chen2004nanotech,Kim2001prb,Li2003nanolett,Li2006prb}.
By applying a transverse electric field,
the band gap of semiconducting CNTs can be closed 
\cite{O'Keefe2002apl,Son2005prl,Chen2004nanotech,Kim2001prb}
and the linear dispersion of metallic CNTs around Fermi level can be bent
 \cite{Li2003nanolett,Li2006prb}.

Inspired by the rich physics and functionalities of CNT
and GNR, researchers have studied nanotubes and nanoribbons made of hexagonal
boron-nitride (BN) sheets,  which are an inorganic analogue of
Gr \cite{Novoselov2005pnas}.
The giant Stark effects have been predicted in BN
nanotubes \cite{Khoo2004prb} and BN nanoribbons 
\cite{Zhang2008prb,Barone2008nanolett,Park2008nanolett}.
For BN nanotubes and nanoribbons, the transverse electric field
shifts the lowest $\sigma^\ast$-band downward, which is often called the
nearly free electron state, and the nanotubes and nanoribbons can be
metallic above a critical field 
\cite{Khoo2004prb,Barone2008nanolett,Park2008nanolett,Zhang2008prb}.
The giant Stark effect of BN nanotubes was experimentally observed by
Ishigami {\it et al}.\ \cite{Ishigami2005prl}.

\begin{figure}[b!]
\centering
\includegraphics[width=7cm]{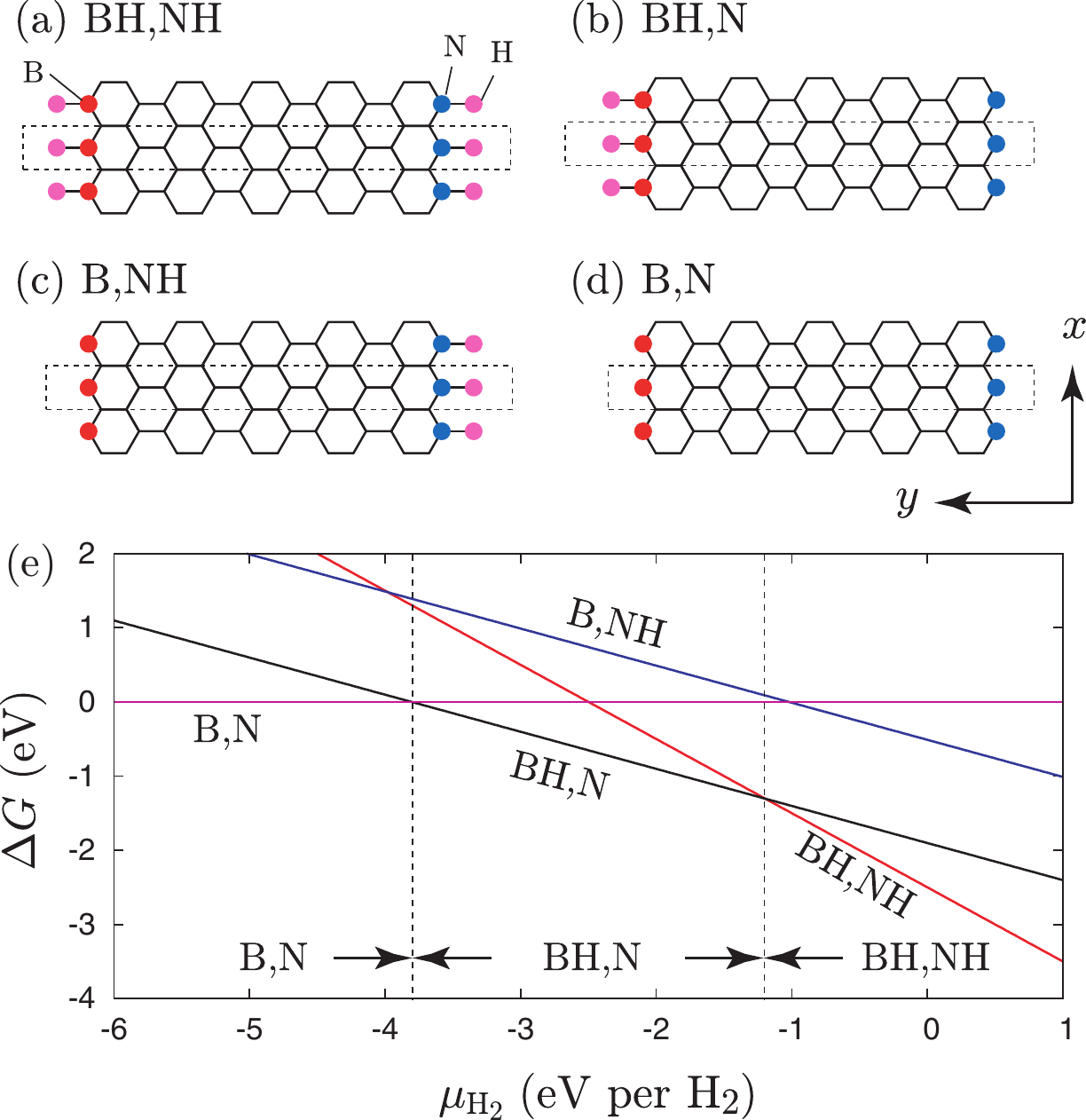}
\caption{(color online)
Schematic illustrations of B-C-N nanoribbons 
with $N=10$ for BH,NH (a), BN,N (b), B,NH (c), and B,N (d).
The dotted rectangles indicate unit cells.
In Panel (e), the difference in the Gibbs free energy, $\Delta G$, is
measured from the Gibbs free energy of the B,N structure
as a function of the chemical potential of H$_2$ gas.
}
\label{fg:Structures}
\end{figure}

On the other hand, Ci {\it et al}. recently reported the synthesis
of hybridized BN and Gr sheets using thermal catalytic chemical vapor
deposition \cite{Ci2010nmat}. Since B and N atoms act as acceptors and donors in Gr, 
respectively, a nanoribbon made of the hybridized BCN
sheet should have rich functionality in controlling electronic and magnetic properties.
Nakamura {\it et al}., He {\it et al}., and Basheer {\it et al}.\
investigated zigzag GNRs with single BN edges and showed
that the ribbons are metallic ferrimagnetic or half-metallic, and that such
properties are realized by the coexistence of extended states and edge-localized states \cite{Nakamura2005prb,He2010apl,Basheer2011njp}.
Kan {\it et al}.\ reported the half-metallic properties of zigzag Gr
and BN hybrid nanoribbons \cite{Kan2008jcp}.

In this letter, we theoretically study the electronic properties of 
the zigzag BCN nanoribbons shown in Figs.\ \ref{fg:Structures} (a)-(d)
using the first-principles calculations.
The outermost C atoms at each edge are replaced by B or N atoms.
Hereafter, we call this structure a B-C-N nanoribbon.
We show that B-C-N nanoribbons are metallic, since several bands cross the Fermi level.
For B-C-N nanoribbons in a rich H$_2$ environment,
the nearly free electron state appears just above the Fermi level
due to the intrinsic giant Stark effect 
arising from the transverse electric dipole moment.
We also show that the intrinsic giant Stark effect 
can be controlled by applying an electric field.
We also discuss the giant Stark effect 
in bilayer BCN nanoribbons.

\begin{figure}[t!]
\centering
\includegraphics[width=7cm]{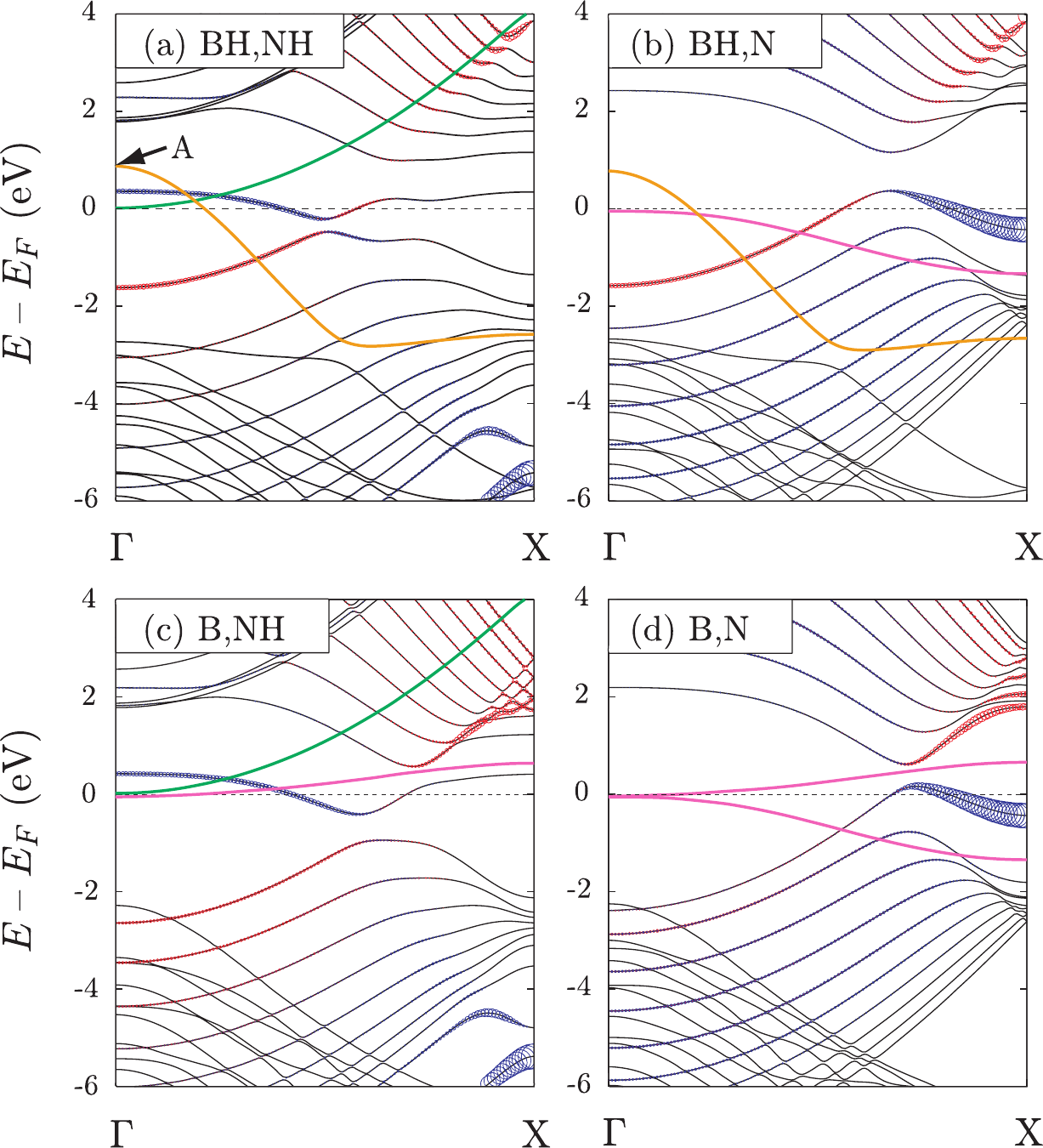}
\caption{(color online)
The band structures of B-C-N nanoribbons with $N=10$ for BH,NH (a), BN,N (b), B,NH (c), and B,N (d).
The projections to $p_z$-orbitals of B and N atoms are respectively indicated by
the red and blue circles whose radii are proportional to the magnitude of the projection.
}
\label{fg:BCN-band}
\end{figure}

\begin{table}[b!]
\centering
\caption{
Calculated results of formation energies at zero temperature $E_{\rm form}$ and dipole moment in the $y$ direction $d_y$.
}
\begin{tabular}{c|cccc}
\hline \hline 
& BH,NH & BH,N & B,NH & B,N \\
\hline 
$E_{\rm form}$ (eV) &\quad $4.542$ \quad &\quad $5.139$ \quad &\quad $6.531$ \quad &\quad $7.042$ \quad \\
$d_y$ ($e$\AA) &$0.67$&$-0.21$&$0.39$&$-0.48$ \\
\hline \hline 
\end{tabular}
\end{table}

We used the projector augmented-wave method \cite{Blochl1994} and the local density
approximation (LDA) \cite{Perdew1981} implemented in the Vienna {\it
ab-initio} simulation package {\scriptsize (VASP)} code
\cite{Kresse1996prb,Kresse1996cms}. 
As shown in Figs.\ \ref{fg:BCN-band} (a)-(d), the B-C-N nanoribbon is located
in the $x-y$ plane and runs along the $x$-axis.
Let $N$ be the number of zigzag lines.
The cell size in the $x$-direction was chosen to be $a=2.446$\AA\,
which is the lattice constant of free-standing Gr optimized in LDA.  
We imposed a vacuum region about 12\AA\ thick in the $y$ and $z$ directions.
The geometry of the nanoribbons was fully optimized 
until the residual force fell below $10^{-3}$ eV/\AA.
The dipole correction in the $y$-direction is used to
exclude spurious dipole interactions between periodic images \cite{Neugebauer1992}.
The cutoff energy of the plane-wave basis was chosen to be
400 eV, and the $k$-points sampling was chosen to be $16\times1\times1$
with a Monkhorst-Pack mesh.

\begin{figure}[t!]
\centering
\includegraphics[width=7cm]{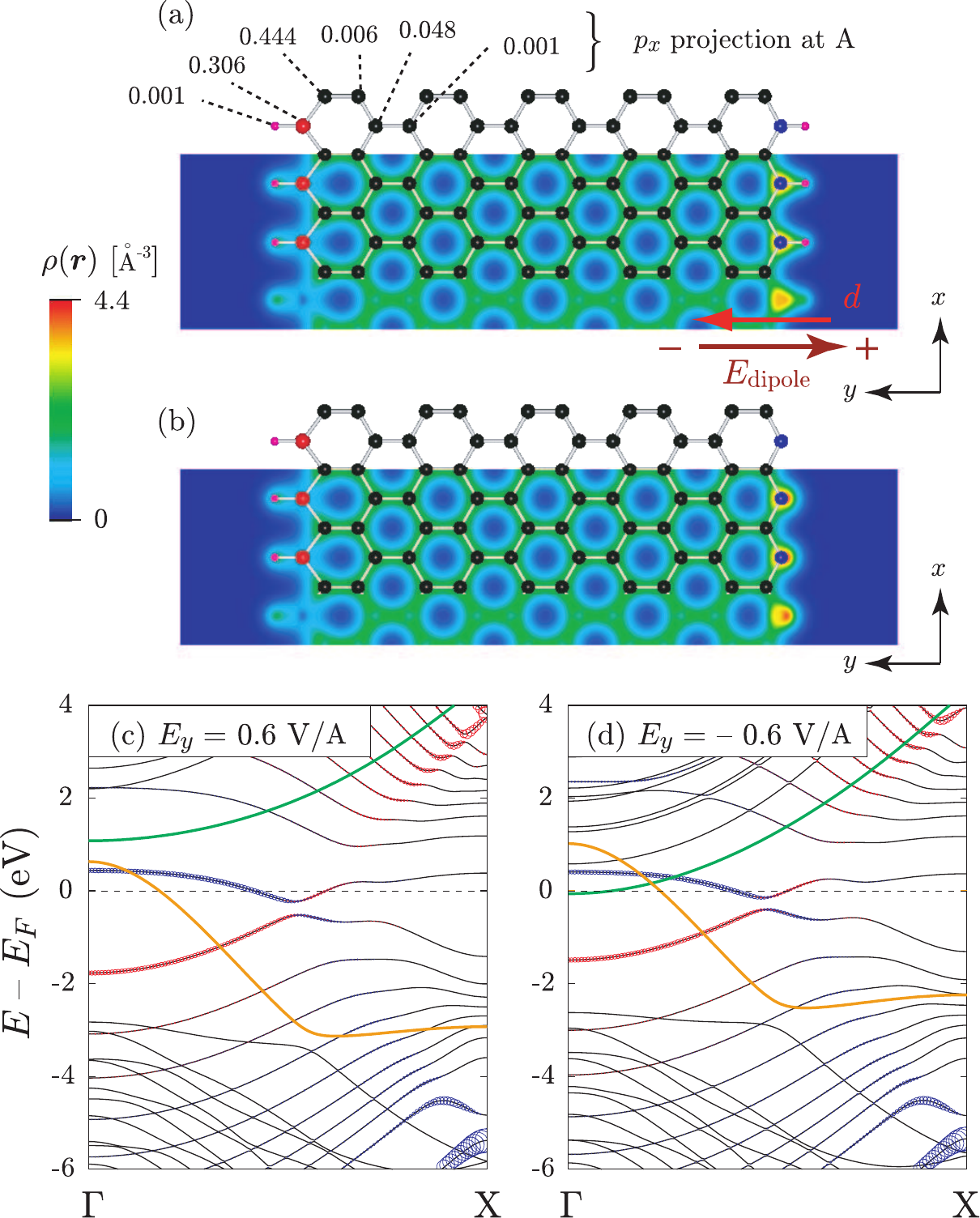}
\caption{(color online)
(a) Optimized structure and electronic density of 
B-C-N nanoribbon with $N=10$ for the BH,NH structure.
(b) The same plot for the BH,N structure.
The magnitude of projection to the $p_x$-orbitals 
of state A, which is indicated in Fig.\ \ref{fg:BCN-band} (a),
is shown at the top of Panel (a).
(c) The band structure of the B-C-N nanoribbon with $N=10$ 
for the BH,NH structure under $E_y=0.6$ V/\AA.
(d) The same plot for $E_y=-0.6$ V/\AA.
The overlap between the $\sigma$-bands can be controlled by $E_y$,
while the $\pi$-bands are not sensitive to $E_y$.
}
\label{fg:Ey-band-N10}
\end{figure}

\begin{figure*}[t!]
\centering
\begin{tabular}{cc}
\begin{minipage}{0.65\hsize}
\includegraphics[width=10.5cm]{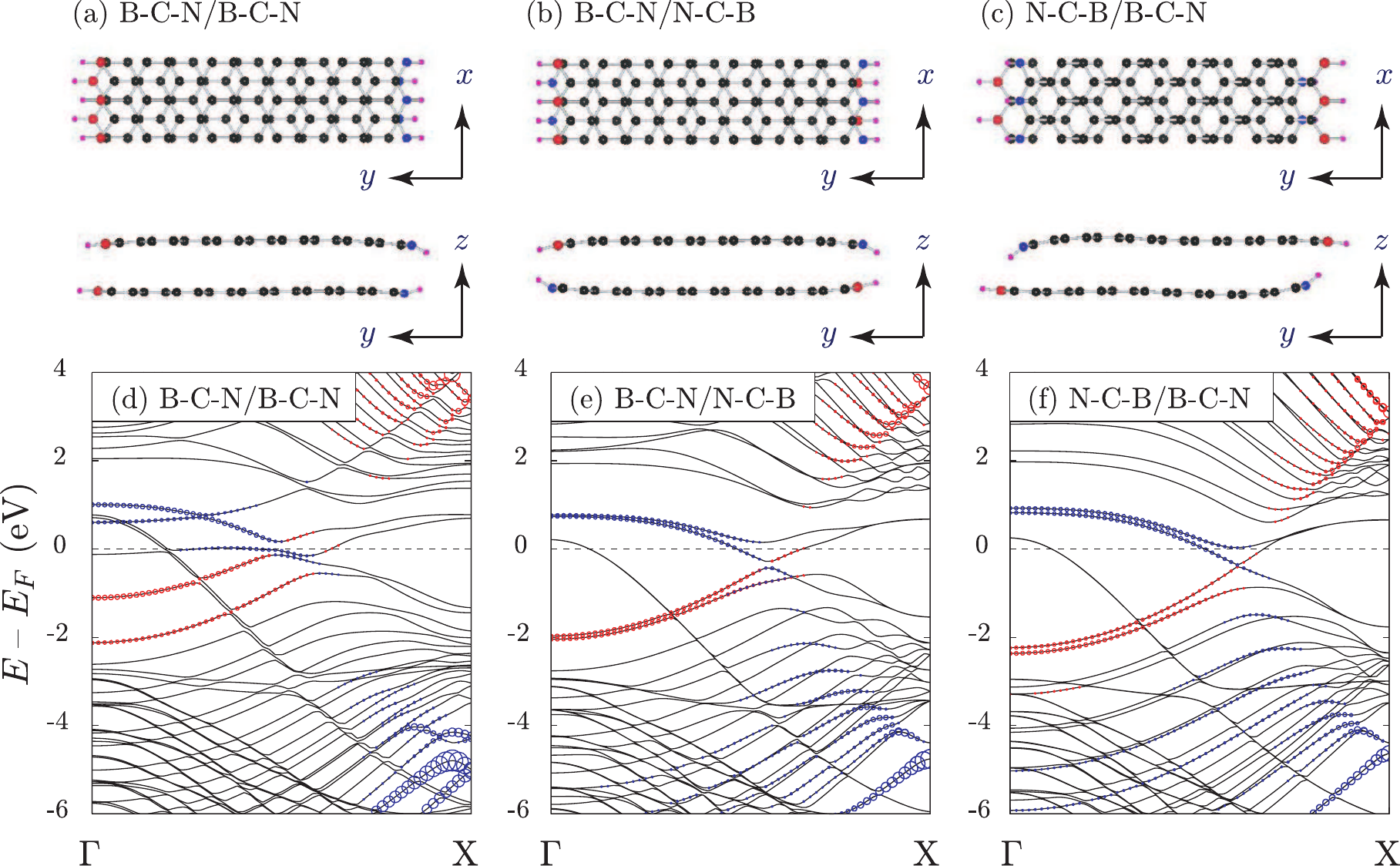}
\end{minipage}
&
\begin{minipage}{0.30\hsize}
\caption{(color online)
(a)-(c) Optimized structures of bilayer B-C-N nanoribbons with different stacking orders
whose edges are terminated hydrogen atoms.
The AB stacking was broken.
(d)-(f) The band structures of bilayer B-C-N nanoribbons 
corresponding to the structures shown in (a)-(c).
The giant Stark effect remains for the B-C-N/B-C-N structure 
but disappears for the B-C-N/N-C-B and N-C-B/B-C-N structures.
}
\label{fg:BCN-bilayer}
\end{minipage}
\end{tabular}
\end{figure*}

First, we discuss the stability of H-terminated B-C-N nanoribbons.
After the geometry is optimized, we obtained perfectly flat nanoribbons independent of H termination.
The formation energy, $E_{\rm form}$, of a B-C-N nanoribbon at zero temperature is defined as 
\begin{equation}
E_{\rm form} = E_{\rm B\mbox{-}C\mbox{-}N} - E_{\rm BN} 
 - \frac{N_{\rm C}}{2}E_{\rm Gr} - \frac{N_{\rm H}}{2}E_{\rm H_2},
\end{equation}
where $E_{\rm B\mbox{-}C\mbox{-}N}$, $E_{\rm BN}$, $E_{\rm Gr}$, and $E_{\rm H_2}$ 
are the total energies of the B-C-N nanoribbon, BN sheet, free-standing Gr, and H$_2$ molecules, respectively. 
Here $N_{\rm C}$ and $N_{\rm H}$ are the numbers of C and H atoms in the unit cell, respectively.
Table 1 summarizes the formation energies of B-C-N nanoribbons with $N=10$ 
for BH,NH, BN,N, B,NH, and B,N.
We found that the BH,NH nanoribbon is stable at zero temperature.
In this comparison, however, the effects of environment, i.e., 
the temperature and pressure of H$_2$ gas, are absent \cite{Wassmann2008prl}.
Stability at a finite temperature can be discussed by means of 
the Gibbs free energy, $G$, which is defined as $G=E_{\rm form}-N_{\rm H}\mu_{{\rm H}_2}/2$, 
where $\mu_{{\rm H}_2}$ is the chemical potential of H$_2$ \cite{Wassmann2008prl,Wang2010prb}.
Figure \ref{fg:Structures} (e) shows the difference in the Gibbs free energy, 
$\Delta G$, measured from the Gibbs free energy of the B,N structure
as a function of the chemical potential of H$_2$ gas.
We found that the BH,NH nanoribbon is the most stable in a H$_2$-rich environment 
and that the B,N nanoribbon is the most stable in a H$_2$-poor environment.
The BH,N nanoribbon becomes stablest in the intermediate environment, 
and the B,NH nanoribbon cannot be realized under practical conditions.

Figure \ref{fg:BCN-band} shows the band structures of B-C-N nanoribbons 
with $N=10$ for BH,NH (a), BH,N (b), B,NH (c), and B,N (d).
The projections to $p_z$-orbitals of B and N atoms are respectively
indicated by the red and blue circles whose radii are
proportional to the magnitude of the projection.
The highest $\sigma$-band and the lowest $\sigma^\ast$-band are 
indicated by the orange and green lines, respectively.
The purple lines correspond to the dangling bonds of B or N atoms, 
which consist mainly of $p_y$-orbitals.

As shown in Fig.\ \ref{fg:BCN-band} (a), the BH,NH structure  
is metallic, since several bands cross the Fermi level.
The magnitude of projection to the $p_x$-orbitals of the highest $\sigma$-band 
at the $\Gamma$ point, which is labeled as A in Fig.\ \ref{fg:BCN-band} (a), 
is shown in the top Panel of Fig.\ \ref{fg:Ey-band-N10} (a).
The projections to the other orbitals, such as $s$ and $p_y$, are zero.
Since the projection to the $p_x$-orbital takes a large value 
along the outermost zigzag line of the right side edge consisting of B atoms,
the corresponding state is localized along the zigzag line, 
i.e., the current carrying edge state.
The bottom of the lowest $\sigma^\ast$-band, indicated by the green curve, 
is located just above the Fermi level.
Since the effective mass of the lowest $\sigma^\ast$ band
is 1.06 $m_0$ with the electron mass in vacuum $m_0$,
this band corresponds to the so-called nearly free electron state 
\cite{Khoo2004prb,Barone2008nanolett,Park2008nanolett,Zhang2008prb}.
The nearly free electron state appears just outside 
of the left side edge consisting of N atoms
similar to that  induced by the giant Stark effect in
BN nanotubes and BN nanoribbons under an external electric field.
Despite the absence of an electric field,
the giant Stark effect seems to take place in the BH,NH structure as shown
in Fig.\ \ref{fg:BCN-band} (a).

The band structure of the BH,N structure is shown in Fig.\ \ref{fg:BCN-band} (b).
We found hole pockets around the $\Gamma$ point and $k\sim 2\pi/3a$.
Around the $\Gamma$ point, the highest $\sigma$-band crosses the Fermi level, 
which can be regarded as the current carrying edge state discussed above.
The purple curve just below the Fermi level corresponds to the dangling bond state of the N atom. 
We did not observe the nearly free electron state for the BH,N structure.
Figure \ref{fg:BCN-band} (c) shows the band structure of the B,NH structure,
where we did not observe the current carrying edge state
but did observe the nearly free electron state.
The purple curve just above the Fermi level corresponds to the dangling bond state of the B atom. 
As shown in Fig.\ \ref{fg:BCN-band} (d), 
$\pi$ and $\pi^\ast$-bands of the B,N structure are quite similar to 
those calculated within tight-binding models \cite{Harigaya2012}.
The purple curves correspond to the dangling bond states of B and N atoms.
From the band structures shown in Figs.\ \ref{fg:BCN-band} (a)-(d),
one can clearly see that the hydrogenation leads 
not only to saturation of the dangling bonds 
but also to considerable changes in the whole band structure.

As mentioned above, the appearance of the nearly free electron state 
reminds us of the giant Stark effect in BN nanotubes and BN nanoribbons
under an external electric field.
However, the nearly free electron state of B-C-N nanoribbons 
appears in the absence of the external electric field, 
which implies the existence of the intrinsic giant Stark effect
due to the internal electric field, i.e., the dipole moment.
Actually, the hydrogenated B-C-N nanoribbons have 
large dipole moments in the transverse direction
due to the large difference in 
electro-negativities among B, C, N, and H atoms.
The electro-negativities of B, C, N, and H atoms 
are 2.04, 2.55, 3.04, and 2.20, respectively.

Calculated dipole moments in the $y$-direction 
per unit cell, $d_y$, are summarized in Table 1.
The value and sign of $d_y$ depend strongly on the hydrogenation pattern.
The electronic density distribution of the BH,NH and BH,N structures 
are shown by the color plot in Figs.\ \ref{fg:Ey-band-N10} (a) and (b), respectively.
From these figures, we found that the hydrogenation of the N atom leads 
to the formation of a large dipole moment in the $+y$-direction 
along the N side edge of the BH,NH structure.
The electric field due to the dipole moment is in the $-y$-direction.

Figures \ref{fg:Ey-band-N10} (c) and (d) show the effects of a transverse electric
field, $E_y$, on the band structure with $N=10$ for the BH,NH structure.
Applying the external electric field in 
the $+y$-direction to reduce the giant Stark effect,
the nearly free electron state is shifted upward, 
as shown in Fig.\ \ref{fg:Ey-band-N10} (c).
In contrast, the application of the external electric field in the $-y$-direction
shifts the nearly free electron states  downward,
as shown in Fig.\ \ref{fg:Ey-band-N10} (d).
Therefore, we conclude that the nearly free electron state in the B-C-N nanoribbon
is caused by the intrinsic giant Stark effect
due to the internal electric field of a dipole moment.

Let us move on to the bilayer B-C-N nanoribbons
where the intrinsic giant Stark effect will depend on the stacking order.
Figures \ref{fg:BCN-bilayer} (a)-(c) show the top and side views of 
optimized structures of the bilayer B-C-N nanoribbons.
We considered three different stacking orders: 
B-C-N/B-C-N (a), B-C-N/N-C-B (b), and N-C-B/B-C-N (c).
The difference between the B-C-N/N-C-B and N-C-B/B-C-N structures is
whether the outermost atom is N (b) or B (c).
While we started the geometry optimization from the 
so-called AB-stacking structure, the AB-stacking was broken,
as shown in the top view of Figs.\ \ref{fg:BCN-bilayer} (a)-(c).
The dipole moment of the B-C-N/B-C-N structure is $d_y=0.66$ e\AA\, 
which induces the intrinsic giant Stark effect as shown in Fig.\
\ref{fg:BCN-bilayer} (d).
For the B-C-N/N-C-B and N-C-B/B-C-N structures, the dipole moments 
of the top and bottom layers cancel each other out, and
the nearly free electron state around the Fermi level disappears 
as shown in Figs.\ \ref{fg:BCN-bilayer}(e) and (f).

In summary, we theoretically studied the electronic properties of BCN
nanoribbons with zigzag edges where the outermost C atoms are uniformly 
replaced by B and N atoms using the first-principles calculations.
BCN nanoribbons are metallic, since several bands cross the Fermi level.
For BCN nanoribbons in a H$_2$-rich environment,
the nearly free electron state appears just above the Fermi level because of the intrinsic giant Stark effect due to the internal
electric field of the transverse dipole moment.
The hydrogenation of N atoms plays a decisive role in the appearance of the giant Stark effect in BCN nanoribbons. 
The strength of the intrinsic giant Stark effect 
can be controlled by applying an external electric field.
With a decrease in the chemical potential of H$_2$ gas, 
the giant Stark effect disappears due to the desorption of H atoms from N atoms.
We also investigated the electronic properties of bilayer BCN nanoribbons 
and showed that the appearance of the giant Stark effect depends on
the stacking order.

{\it Acknowledgments} - 
The authors acknowledge Y.\ Shimoi, H.\ Arai, T.\ Nakanishi, 
K.\ Wakabayashi, S.\ Dutta, and M.\ B\"urkle for valuable discussions.
This research was supported by the International Joint Work Program of 
Daeduck Innopolis under the Ministry of Knowledge Economy (MKE) of the Korean Government.

\end{document}